\documentclass[8.5pt,twoside,twocolumn]{article}
\usepackage[super,sort&compress,comma]{natbib} 
\usepackage{mhchem}
\usepackage{times,mathptmx}
\usepackage{sectsty}
\usepackage{balance} 

\usepackage{mciteplus} 

\usepackage{graphicx} 
\usepackage{lastpage}
\usepackage[format=plain,justification=raggedright,singlelinecheck=false,font=small,labelfont=bf,labelsep=space]{caption} 
\usepackage{fancyhdr}
\usepackage{amsmath}

\begin{document}

\def\cal#1{\mathcal{#1}}
\def\eqq#1{Eq.~(\ref{#1})}
\def\eqs#1{Eqns.~(\ref{#1})}
\def\eq#1{(\ref{#1})}

\def\f#1{Fig.~\ref{#1}}
\def\ff#1{Figure~\ref{#1}}
\def\fs#1{Figs.~\ref{#1}}

\def\s#1{Section~\ref{#1}}

\def\c#1{~\cite{#1}}

\def\beq{\begin{equation}}
\def\eeq{\end{equation}}
\def\bea{\begin{eqnarray}}
\def\eea{\end{eqnarray}}

\twocolumn[
  \begin{@twocolumnfalse}
\noindent\LARGE{\textbf{Crystallization and arrest mechanisms of model colloids}}
\vspace{0.6cm}

\noindent\large{\textbf{Thomas K. Haxton,$^{\ast}$\textit{$^{a}$} Lester O. Hedges,\textit{$^{a,b}$} and
Stephen Whitelam$^{\ast}$\textit{$^{a}$}}}\vspace{0.5cm}

\noindent\textit{\small{\textbf{Received Xth XXXXXXXXXX 20XX, Accepted Xth XXXXXXXXX 20XX\newline
First published on the web Xth XXXXXXXXXX 200X}}}

\noindent \textbf{\small{DOI: 10.1039/b000000x}}
\vspace{0.6cm}

\noindent \normalsize{
We performed dynamic simulations of spheres 
with short-range attractive interactions
for many values of interaction
strength and range. 
Fast crystallization occurs
in a localized region of this 
parameter space,
but the character of crystallization pathways is not uniform within this region.
Pathways range from 
one-step, 
in which a crystal nucleates directly from a gas, 
to
two-step, 
in which substantial liquid-like clusters form and only subsequently become crystalline. Crystallization can fail because of slow nucleation from either gas or liquid, or because of dynamic arrest caused by strong interactions. Arrested states are characterized by the formation of networks of face-sharing tetrahedra that can be detected by a local common neighbor analysis.
}
\vspace{0.5cm}
 \end{@twocolumnfalse}
  ]



\footnotetext{$^{a}$Molecular Foundry, Lawrence Berkeley National Laboratory, Berkeley, CA 94720, United States}
\footnotetext{$^{b}$Department of Physics, University of Bath, Bath, BA2 7AY, United Kingdom}

Colloidal crystallization is of considerable interest because of the value of colloidal assemblies to technology~\cite{dinsmore1998self,Talapin2010} and the value of colloidal dispersions as model systems~\cite{Anderson2002, Lu2013, Gao2011}. Colloidal particles can be made of controlled size, interaction strength, and interaction range, 
making them useful models for exploring
the thermodynamic and kinetic factors that lead to the assembly of equilibrium and nonequilibrium condensed states of matter~\cite{leunissen2005ionic,ivlev2012complex,mao1995depletion}.

A defining feature of many colloidal suspensions is that their interactions can be short ranged compared to the nm to $\mu$m size of the colloidal particles.  For example, van der Waals interaction, depletion interactions~\cite{Tuinier2003}, and DNA base-pairing interactions~\cite{Gao2011, Biancaniello2005, Park2008, Nykypanchuk2008} used to promote colloidal crystallization typically act over a range of distances small compared to the particle size.  As a result, colloidal suspensions can exhibit phase behavior and assembly kinetics not typically seen in atomic or molecular systems~\cite{asherie1996phase, tenWolde1997, lutsko2006ted, Soga1999, Costa2002, Lomakin2003, Fortini2008}.

\begin{figure*}
\includegraphics[width=\textwidth]{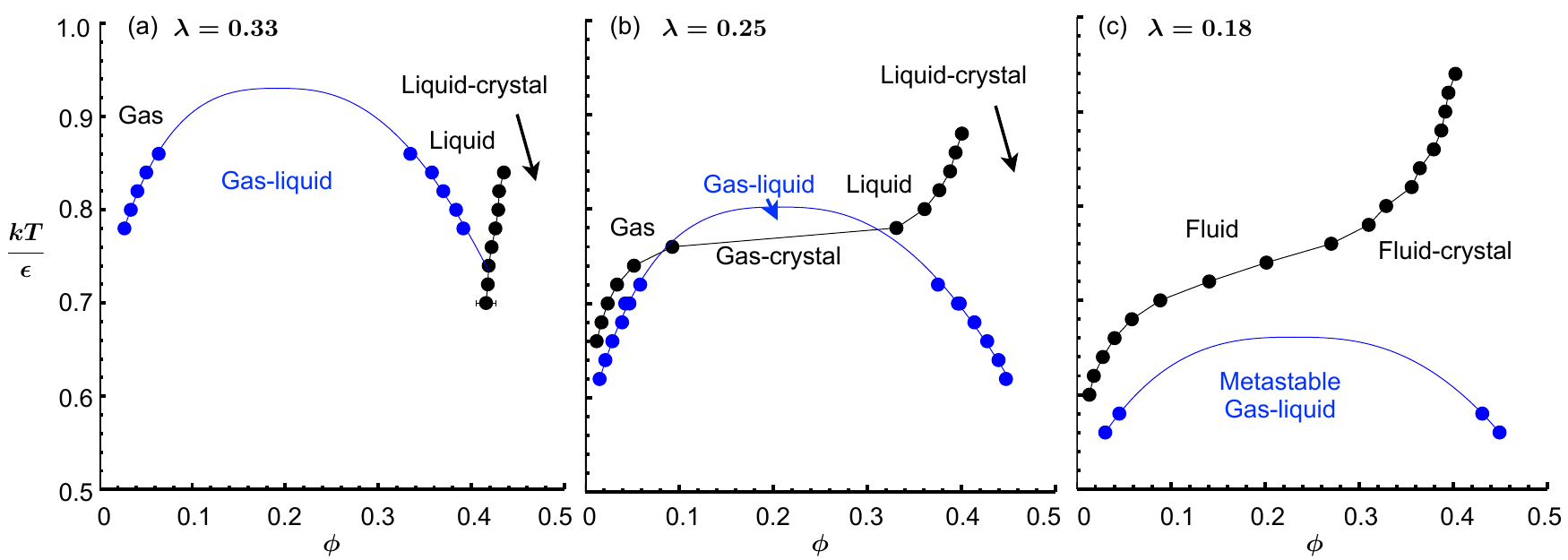}%
\caption{Phase diagrams of hard spheres of radius $R_0$ with attractive square-well interactions of range $2(1+\lambda) R_0$, in the plane of density (hard-core packing fraction $\phi$) and temperature (scaled interaction strength $k_{\rm B} T/\epsilon$). Interaction ranges are (a) $\lambda=0.33$, (b) $\lambda=0.25$, and (c) $\lambda=0.18$.  Blue circles represent densities of coexisting fluids, calculated using Gibbs ensemble Monte Carlo simulations, and blue curves are fits to the coexistence curve for systems in the Ising universality class. Black circles represent densities of fluids coexisting with face-centered-cubic (fcc) crystals, calculated using direct coexistence simulations.  Crystal densities, to the right of these plots, are not shown.}
\label{phasediagrams}
\end{figure*}

A minimal and well-studied model of a colloidal dispersion is a collection of spheres of radius $R_0$ interacting via the `square-well' potential~\cite{Alder1972, Young1980, Liu2005, Pagan2005, asherie1996phase}
\begin{equation}
U(r_{ij})=\left\{\begin{array}{cc}
\infty & r_{ij}\le 2R_0 \\
-\epsilon & 2R_0< r_{ij} \le 2(1+\lambda) R_0 \\
0 & r_{ij}>2(1+\lambda) R_0, \\
\end{array}\right.
\end{equation}
where $r_{ij}$ is the distance between the centers of particles $i$ and $j$. In this model the solvent in which particles are dispersed is not represented explicitly. Square-well spheres display the well-known phase behavior of particles with isotropic attractions~\cite{tenWolde1997,lutsko2006ted,charbonneau2007systematic}, 
summarized in Fig.~\ref{phasediagrams}. When the interaction range $\lambda$ is large (Fig.~\ref{phasediagrams} (a)), the square-well system exhibits sequential phase transitions from gas to liquid to crystal as temperature (i.e. the combination $k_{\rm B}T/\epsilon$) decreases. (Here ``gas" refers to a dilute suspension of colloids and ``liquid" to a concentrated suspension.)  As  the interaction range decreases, the gas-liquid coexistence curve decreases in temperature (Fig.~\ref{phasediagrams} (b)), eventually becoming metastable with respect to fluid-crystal coexistence (Fig.~\ref{phasediagrams} (c)).

Many studies have shown that such metastable liquid-gas phase separation can play a crucial role in colloidal crystallization. Free-energy calculations~\cite{tenWolde1997, lutsko2006ted}, and dynamic simulations~\cite{Soga1999, Costa2002, charbonneau2007systematic, Fortini2008, klotsa2011predicting} of short-range attractive spheres show that the metastable liquid can promote a 
two-step
crystallization pathway in which colloids coalesce into liquid droplets from which crystals nucleate. Two-step pathways have been observed experimentally in colloidal particles confined in two dimensions~\cite{Hobbie1998, Zhang2007, Savage2009}, in DNA-tethered nanoparticles~\cite{Macfarlane2009}, and in proteins~\cite{Galkin2000,Galkin2001}, as well as in simulations of DNA-tethered nanoparticles~\cite{Dai2010}.

Colloidal liquid-gas phase separation also plays an important role in the formation of (physical) gels at deep supercooling. Gels are nonequilibrium, disordered networks of particles with solid-like mechanical properties that result from their percolating structures. Gelation occurs because strong inter-particle bonding causes particles within aggregates to rearrange too slowly to allow equilibration on observed timescales. In the deeply supercooled spinodal regime, rapid liquid-gas phase separation can cause the formation of extended, non-compact colloidal aggregates which fail to relax into compact colloidal droplets~\cite{Anderson2002, Zaccarelli2007}. Many experimental studies of polymeric colloidal particles possessing depletion attractions have found gelation to occur in preference to crystallization~\cite{Campbell2005, Sciortino2005b} (potentially exacerbated by effects of polydispersity~\cite{Sollich2010, Zhang2013}). Microscopic analysis of the colloid-colloid interaction networks formed during gelation shows gelation to be characterized by certain (overlapping) locally-favored motifs: gels in long-range repulsive colloids consist of networks of face-sharing tetrahedra (maximally bonded clusters of four particles)~\cite{Campbell2005, Sciortino2005b}, while gels of short-range-attractive spheres consist of networks of maximally bonded clusters of various sizes~\cite{Royall2008}.

Recently, several authors have investigated the generality of crystallization and gelation mechanisms by characterizing colloid dynamics across broad sections of parameter space.  Macfarlane \textit{et al}.~showed in experiments that DNA-linked nanoparticle crystallization occurs for each nanoparticle size only within a limited range of DNA lengths.
Short lengths resulted in effective interaction ranges smaller than the nanoparticles' polydispersity, disfavoring the crystal thermodynamically, while large lengths inhibited kinetics~\cite{Macfarlane2010}.  

Several authors have performed Monte Carlo, molecular dynamics, or Brownian dynamics simulations of spheres with short-range attractive interactions to investigate how assembly mechanism and product depend on interaction strength and/or concentration.  Multiple studies have shown that crystallization occurs via two-step nucleation for a window of temperatures below the metastable liquid-gas transition, with deeper temperature quenches leading to gelation~\cite{Soga1999, Costa2002, charbonneau2007systematic}.  Extending these studies to multiple concentrations, Fortini \textit{et al.}~showed that crystallization coincides with the metastable liquid-gas transition temperature at concentrations below the liquid-gas critical concentration but occurs also at higher temperatures at supercritical concentrations~\cite{Fortini2008}.  Performing single-particle Monte Carlo simulations of square-well spheres at a single packing fraction of $\phi=0.04$, Klotsa and Jack found an exception to the two-step rule: at a temperature near the metastable liquid-gas transition, they found that crystallization can proceed via a one-step pathway without significant formation of amorphous clusters~\cite{klotsa2011predicting}.


Complementing and extending these studies, we describe in this paper the self-assembly dynamics of square-well-attractive spheres over a broad spectrum of interaction strengths and ranges.  We simulated sphere dynamics using the virtual-move Monte Carlo algorithm~\cite{Whitelam2007, Whitelam2009role, Whitelam2011approximating, Haxton2012slayer} parameterized so that colloidal clusters diffuse at rates agreeing with Stokes' law.  Consistent with the previously mentioned studies at fixed interaction range, we find that efficient crystallization occurs in a localized region of parameter space, with a high-temperature boundary associated with the metastable liquid-gas transition.  However, we find that the character of crystallization pathway varies within the region of efficient crystallization.  Near the high-temperature boundary, crystallization proceeds along a one-step pathway via nucleation from the gas.  Further below this boundary, crystallization proceeds along a two-step pathway via the formation of liquid-like clusters from which crystals subsequently nucleate.  We find that poor crystallization at low temperature is characterized by the formation of networks of face-sharing tetrahedra that can be detected by a local common neighbor analysis~\cite{Honeycutt1987}.

It is important to note that the square-well model we have studied neglects features of real colloidal particles that may lead to complexity beyond that discussed here. The pairwise nature of the square-well interaction cannot capture collective properties of counterions, depletants, or polymer coats that mediate multi-body interactions between colloidal particles.  For example, counterion entropy can favor gelation over crystallization in a way that cannot be modeled at a pairwise level~\cite{Schmit2011}, and the entropy of mobile linkers in the dilute-linker limit can favor the liquid state to the point of removing the triple point from Fig.~\ref{phasediagrams} (a)~\cite{Martinez-Veracoechea2011}. The square-well model also treats solvent in an implicit manner; explicitly accounting for solvent and the long-ranged hydrodynamics it mediates may be important for colloidal crystallization under certain conditions\c{tanaka2000simulation,cates2004simulating}.

\section{Methods}

\subsection{Structure characterization}
\label{struct}

\begin{figure}[t]
\includegraphics[width=0.5\textwidth]{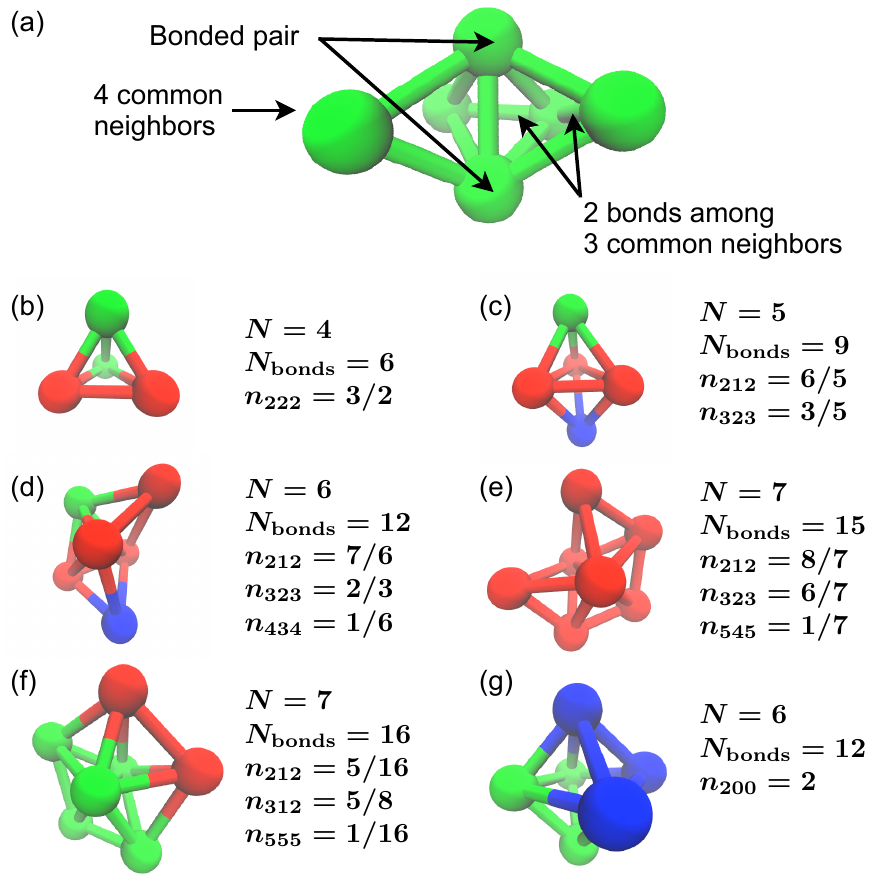}%
\caption{(a) Illustration of the common neighbor analysis for a bonded pair of spheres with a 423 common neighbor environment: the bonded pair shares four common neighbors, there are two bonds among those common neighbors, and three common neighbors participate in those two bonds.  (b-g) Low-energy bond topologies for clusters of size $N\le 7$.  Linear polytetrahedral networks (b-e) and a closed 5-loop of face sharing tetrahedral (f) exhibit nonzero values of $n_{212}, n_{323}, n_{434}, n_{545}$, and/or $n_{555}$.  The octahedron (g) is the only maximally bonded cluster for $N \le 13$ that is not composed of face-sharing tetrahedra.  It has the same number of spheres and bonds as the 3-tetrahedron (d) and a different common neighbor signature, $n_{200}=2$.}
\label{clusters}
\end{figure}

We characterized the dynamics by performing a common neighbor analysis of the network of spheres linked by favorable square-well interactions, similar to the analysis developed by Honeycutt and Andersen~\cite{Honeycutt1987}.  At regular time intervals, we recorded the number of bonded pairs of particles $N_{abc}$ with $a$ common neighbors, $b$ bonds among those common neighbors, and $c$ common neighbors participating in those bonds (see Fig.~\ref{clusters} (a)).  We defined the relative number $n_{abc} \equiv N_{abc}/N$, where $N$ is the number of particles.  This analysis identifies gaseous, liquid, crystalline, and polytetrahedral configurations.  Perfect face-centered cubic and hexagonally close packed crystals exhibit nonzero values only of $n_{423}$ (bulk hexagonal close-packed (hcp)), $n_{424}$ (bulk hcp and fcc), $n_{212}$, and $n_{312}$ (boundaries).  Weakly interacting gases exhibit few bonds and uniformly low values of all common neighbor metrics, while weakly structured liquids exhibit large values of $n_{200}$.  Networks of face-sharing tetrahedra exhibit large values of $n_{323}, n_{434}, n_{545}$, and/or $n_{555}$ (see Fig.~\ref{clusters} (b-f)).  We characterize the crystallinity by the fraction $f_{\rm c}$ of particles participating in at least one 423 or 424 bond.

To quantify the difference in crystallinity between dynamic simulations and simulations begun from fcc or bcc crystals (see below), we use the `distance-to-equilibrium' parameter
\begin{equation}
\Delta_{\rm eq}\equiv\textrm{max}\left(\dfrac{n_{42x}^{\rm fcc}-n_{42x}^{\rm random}}{n_{42x}^{\rm perfect}}, \dfrac{n_{666}^{\rm bcc}-n_{666}^{\rm random}}{n_{666}^{\rm perfect}}\right).
\label{deq}
\end{equation}
Here the superscript `random' describes yield of common neighbor types from dynamic (randomly-initialized) simulations, the superscripts `fcc' and `bcc' describe yields obtained from fcc- and bcc-initialized simulations, respectively, and the superscript `perfect' describes the yield of a given environment in a bulk (perfect) fcc or bcc crystal, i.e. $n_{42x}^{\rm perfect}=6$ and $n_{666}^{\rm perfect}=4$.

\begin{figure}
\begin{centering}
\includegraphics[width=0.5\textwidth]{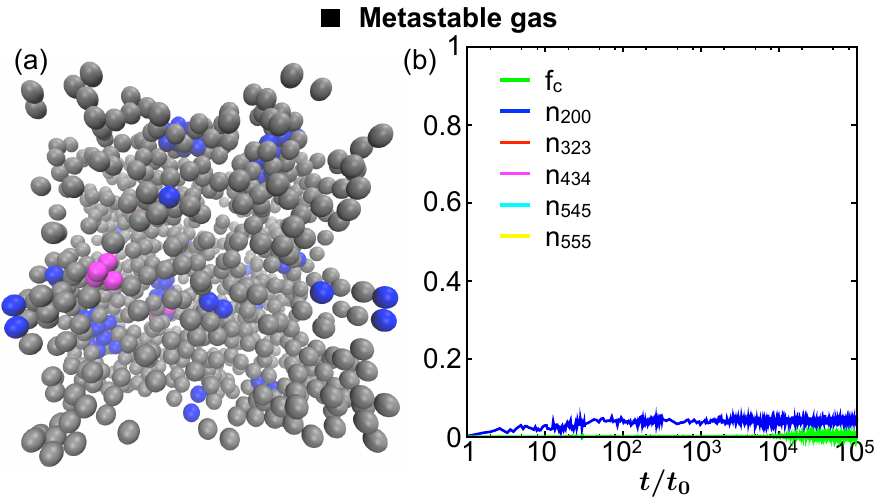}%
\caption{(a) Snapshot from a simulation with $\lambda=0.03$ and $k_{\rm B}T/\epsilon=0.32$. The system remains in the metastable gas phase up to time $t=10^5t_0$. Particle color code is described in \s{struct}. (d) Time series of the crystal yield $f_{\rm c}$ (green), liquid (blue) and polytetrahedral common neighbor metrics.}
\label{metastable}
\end{centering}
\end{figure}

In Figs.~\ref{metastable},~\ref{onestep},~\ref{twosteppathway},~\ref{metastableliquid}, and~\ref{gel} we use the following color code to denote particle environments. Particles participating in crystalline common neighbor environments (423 or 424) are colored green. Particles that do not, and that participate in polytetrahedral  common neighbor environments (323, 434, 545, or 555) are colored red. The remaining particles that participate in in liquid common neighbor environments (200) are colored blue. The remaining particles that participate in other common neighbor environments $abc$ with $a\ge 2$ are colored magenta. The remaining (gas) particles are colored gray.  

\subsection{Thermodynamics}

\begin{figure*}[t]
\includegraphics[width=\textwidth]{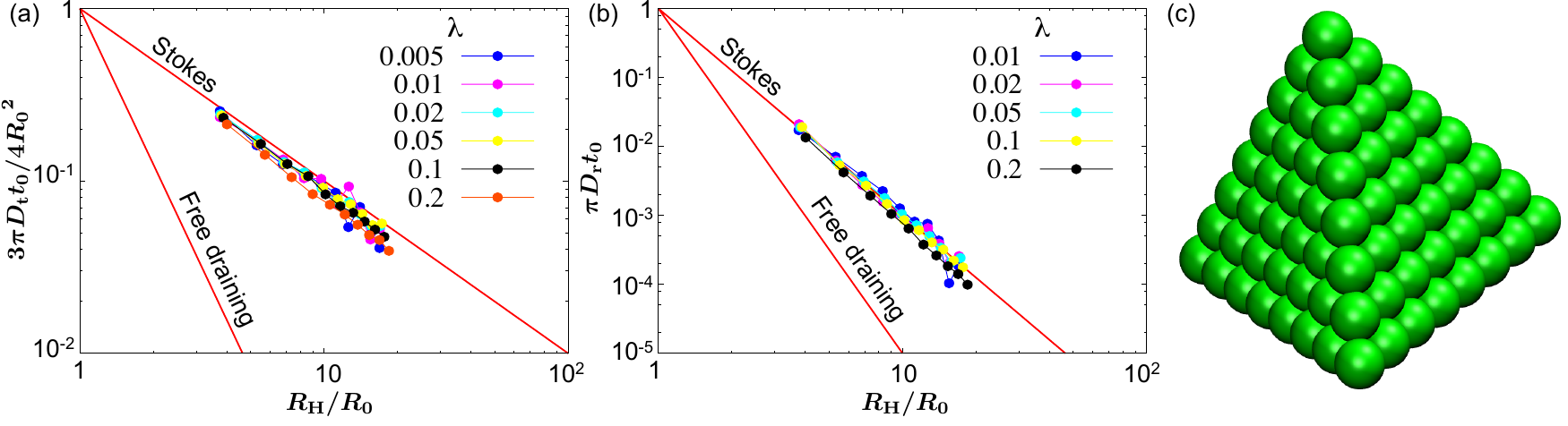}%
\caption{(a) Translational and (b) rotational cluster diffusion constants, as a function of hydrodynamic radius $R_{\rm H}$, for virtual-move Monte Carlo simulations of tetrahedral clusters composed of infinitely attractive square-well spheres. We show results for various interaction ranges $\lambda$ (legend). The clusters used in these calculations were tetrahedral, and ranged in size from 1 to 120 particles; panel (c) shows a cluster of 120 particles. Algorithm parameters were $\Delta_{\rm t}=4\lambda R_0$, $\Delta_{\rm r}=1$, and $p_{\rm t}=0.4(R_0\Delta_r/\Delta_t)^2$ (see text). Cluster diffusion constants approximate the Stokes solutions ($D_{\rm t}\propto {R_{\rm H}}^{-1}$ and $D_{\rm r}\propto {R_{\rm H}}^{-3}$). For comparison, we show also the free-draining solutions ($D_{\rm t}\propto {R_{\rm H}}^{-3}$ and $D_{\rm r}\propto {R_{\rm H}}^{-5}$), which describe substantially slower collective motion.}
\label{diffusion}
\end{figure*}

We tested the thermodynamic stability of finite-size crystals by performing virtual-move Monte Carlo simulations (see below) initialized from compact crystals containing approximately 1000 spheres in a simulation box with an overall hard-core packing fraction of $\phi=0.1$.  We used a 923-particle cuboctahedron for the fcc crystal~\cite{fcccluster} and a 1001-particle cuboctahedron for the bcc crystal~\cite{bcccluster}.
We initialized these crystals with nearest-neighbor distances $d=2(1+\lambda/2)R_0$, placing nearest neighbors in the middle of their interaction range.
We defined the boundary of finite-size fcc stability (Fig.~\ref{yield} (a)) by the contour where $n_{42x}=n_{423}+n_{424}=1$.  (As shown by the relatively sharp decay of $n_{42x}$ in Fig.~\ref{thermo} (a), the location of the boundary is relatively insensitive to choice of threshold.)

We calculated the boundary of stability of bulk fcc crystals by performing single-particle Monte Carlo (SPMC) direct coexistence simulations of 1000 spheres in a slab geometry at various temperatures and interaction ranges, choosing overall packing fractions that allowed sufficient sampling of both fluid and crystal phases.  
We initialized the crystal slabs with nearest-neighbor distances $d=2(1+\lambda/2)R_0$, and we initialized the fluid phases with random configurations without hard-core overlaps.
As shown for example by the black points in Fig.~\ref{phasediagrams}, these simulations allowed us to determine the coexistence concentrations for the fluid phase (gas or liquid, depending on $T$ and $\lambda$) and the crystal phase (not shown).  We defined the boundary of bulk crystal stability at $\phi=0.1$ for each $\lambda$ as the temperature at where the interpolated fluid coexistence concentration (black curves in Fig.~\ref{phasediagrams}) intersects $\phi=0.1$.

We calculated the boundary of stability of the bulk liquid by performing SPMC Gibbs ensemble Monte Carlo simulations of 1000 spheres separated in two boxes that exchange spheres and volume~\cite{Panagiotopoulos1989}.  Analogous to the direct coexistence simulations, we performed the Gibbs ensemble simulations at a range of temperatures for each interaction range, with overall packing fractions chosen to allow sufficient sampling of both phases, and we determined the boundary of liquid (meta)stability at $\phi=0.1$ by interpolating the fluid coexistence curves (see blue points and curves in Fig.~\ref{phasediagrams}).
We initialized both the gas and liquid box with random configurations without hard-core overlaps.

\subsection{Dynamics}

To approximate the overdamped dynamics of strongly-associating particles in solution we used the virtual-move Monte Carlo algorithm~\cite{Whitelam2007} (specifically, the version of the algorithm described in the appendix of Ref.\cite{Whitelam2009role}). Under this algorithm, which satisfies detailed balance, particles move locally according to the gradients of potential energy they experience, and collectively with a rate that can be controlled to a degree by the user. We parameterized the algorithm in a manner similar to that described in Ref.\c{Haxton2012slayer}, in order to ensure that tightly-bound clusters of particles of hydrodynamic radius $R_{\rm H}$ diffused with rates close to those predicted by the Stokes' law,
\begin{equation}
\begin{array}{l}
D_{\rm t}=\dfrac{k_{\rm B}T}{6\pi\eta R_{\rm H}}, \\
D_{\rm r}=\dfrac{k_{\rm B}T}{8\pi\eta {R_{\rm H}}^3}.
\end{array}
\label{Dstokes_main}
\end{equation}
The natural time unit of this motion is then the Brownian time scale
\begin{equation}
t_0= \frac{\eta (2R_0)^3}{k_{\rm B}T},
\label{t0}
\end{equation}
where $\eta$ is the (implicit) solvent viscosity and $k_{\rm B}T$ is the thermal energy. Simulations performed at particular values of $k_{\rm B}T/\epsilon$ and $\lambda$ can therefore be considered to apply to a wide range of absolute particle sizes $R_0$, the latter determining only the value of $t_0$. For example, for spherical colloidal particles with radius $R_0=50$ nm at room temperature ($T=293$ K) in water ($\eta=1.00 \times 10^{-3}$ Pa s), the Brownian timescale is $t_0=2.5 \times 10^{-4}$ s. 

In the Appendix we describe in detail the procedure we used. Briefly, the virtual-move algorithm generates collective Monte Carlo moves by proposing trial `virtual' particle translations or rotations, and probabilistically recruiting neighboring particles to join this motion, in an iterative fashion. The resulting trial move is accepted with a probability ensuring detailed balance. In addition, one is free to attenuate the rate at which collective motion is accepted, by imposing what are effectively kinetic constraints. We chose these constraints in order to enforce \eqq{Dstokes_main}. 

Each Monte Carlo move begins with either a trial translation or a trial rotation, chosen with probability $p_{\rm t}$ and $p_{\rm r}=1-p_{\rm t}$, respectively.
For translations, we randomly selected a particle and translated it randomly within a ball of radius $\Delta_{\rm t}$. 
For rotations, we randomly selected a particle, randomly selected a second particle within the interaction range of the first, and rotated the second particle by an angle, chosen uniformly from the range $(-\Delta_{\rm r}, \Delta_{\rm r})$, around a randomly-oriented axis $\hat{n}$ passing through the center of the first particle. We chose $\Delta_{\rm t}=4\lambda R_0$, $\Delta_{\rm r}=1$, and $p_{\rm t}=0.4(R_0\Delta_r/\Delta_t)^2$. 
As discussed in the Appendix, we found that with this choice of parameters we could enforce \eqq{Dstokes_main} by suppressing the acceptance rate for translation and rotation of a cluster of $N$ particles of hydrodynamic radius $R_{\rm H}$ by factors $N^{-1} R_{\rm H}^{-1}$ and $N^{-1} R_{\rm H}^{-3}$, respectively. We chose to maximize the ratio of rates of internal cluster relaxation to whole-cluster diffusion, by working with the smallest trial displacement $\Delta_{\rm t}$ that is large enough to induce substantial collective motion (i.e. is large enough to ensure that Stokes' law could be maintained). This choice is somewhat arbitrary, and whether it is physically appropriate 
will likely depend on details of the experimental system one wishes to model,
but we note that one has some freedom to influence this ratio if necessary.

In Fig.~\ref{diffusion} we show measured diffusion constants for tetrahedral clusters of between 1 and 120 square-well spheres, in the $k_{\rm B} T/\epsilon\rightarrow 0$ limit. $D_{\rm t}$ and $D_{\rm r}$ approximate the Stokes solutions (Eq.~\ref{Dstokes_main}) over a broad range of cluster sizes and interaction ranges. In this respect our procedure therefore captures an important aspect of solvent-mediated diffusion, without representing solvent explicitly. Note that `long-ranged' hydrodynamic coupling\c{tanaka2000simulation}
is not taken into account by this procedure; to do so, one should represent solvent more explicitly\c{ihle2001stochastic,pooley2005kinetic,sane2009hydrodynamics,cates2004simulating}. Simple implementations of Brownian (Langevin) dynamics integrators, and single-particle Monte Carlo simulations in the limit of zero trial displacement\c{kikuchi1991metropolis}, result instead in the `free-draining' behavior $D_{\rm t}\propto {R_{\rm H}}^{-3}$ and $D_{\rm r}\propto {R_{\rm H}}^{-5}$.  As shown in Fig.~\ref{diffusion} (note the logarithmic scale), such diffusion is significantly slower 
than Stokes' diffusion, even for relatively modest cluster sizes. 

As discussed in the Appendix, our procedure yields a time per Monte Carlo cycle of
\begin{equation}
t_{\rm cycle}=\dfrac{6}{5}\pi p_{\rm t}\lambda^2 t_0,
\end{equation}
where $t_0$ is the Brownian time scale (Eq.~\ref{t0}).  We present results relative to the physical time unit $t_0$. 

We carried out simulations of 1000 square-well spheres, in periodically-replicated cubic simulation boxes, at a hard-core packing fraction of $\phi = 0.1$. We carried out independent simulations for interaction ranges between (and including) the values $\lambda=0.005$ and 1.35, and for temperatures ranging from $k_{\rm B}T/\epsilon=0.06$ to 0.86. We initialized dynamic simulations with random configurations, under the constraint that the particle hard cores could not overlap (equivalent to equilibrium configurations in the $k_{\rm B}T/\epsilon\rightarrow \infty$ limit).  

An open-source C++ library for implementing the virtual-move Monte Carlo algorithm is available at \href{http://vmmc.xyz}{http://vmmc.xyz}~\cite{vmmc}.

\section{Results}

\subsection{Dynamic and thermodynamic phase diagrams}

\begin{figure*}
\includegraphics[width=\textwidth]{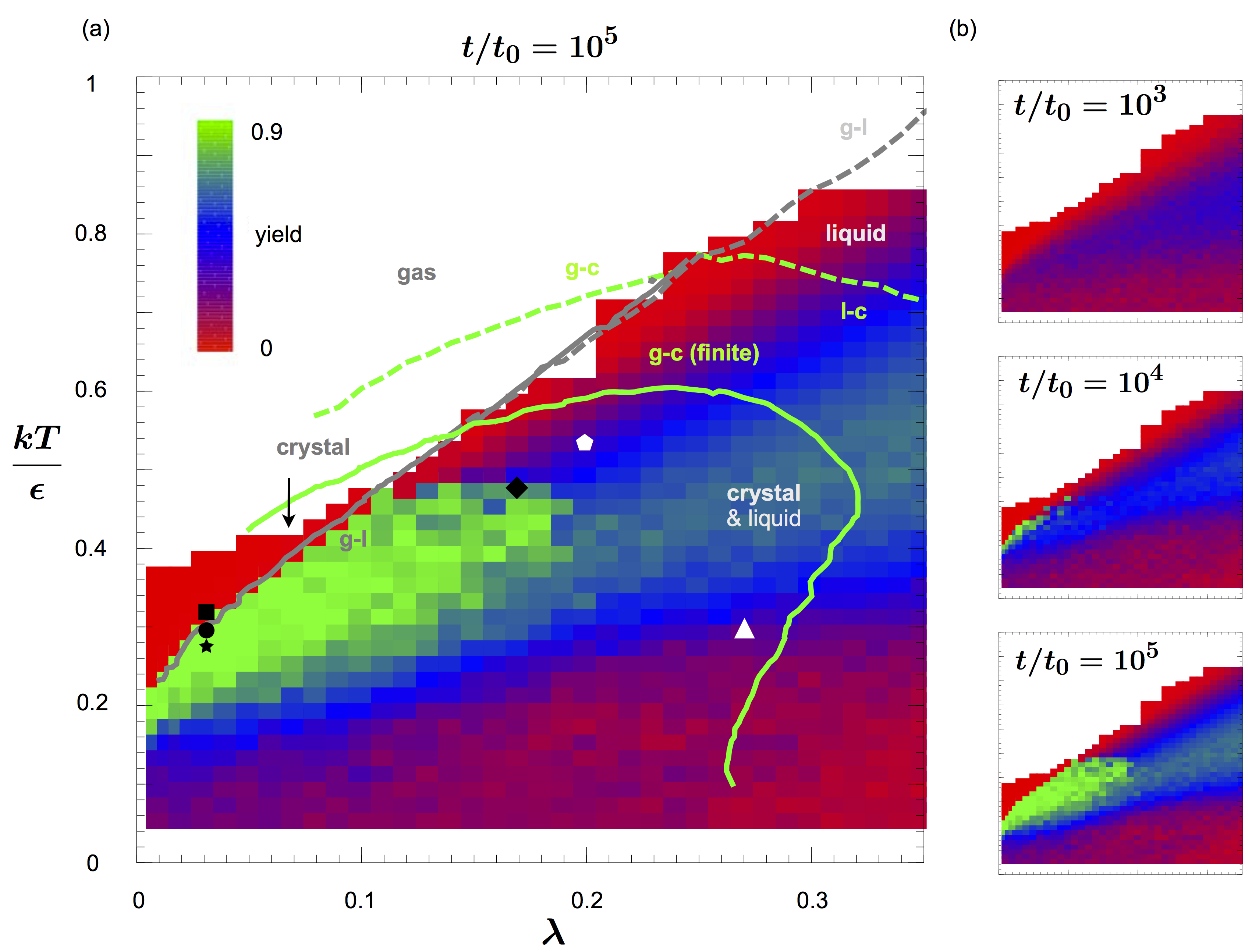}%
\caption{(a) Crystal yield $f_{\rm c}$ from dynamic simulations after $t=10^5t_0$ as a function of interaction range $\lambda$ and temperature $k_{\rm B}T/\epsilon$, for systems of $1000$ square-well spheres at hard-core packing fraction $\phi=0.1$.  The dashed (solid) green curve indicates the boundary of stability of bulk (finite-size) fcc crystals, and the dashed (solid) gray curve indicates the boundary of stability of the bulk (finite-size) liquid (see text for details).   Symbols indicate representative state points for the various dynamic regimes shown in Figs.~\ref{metastable},~\ref{onestep},~\ref{twosteppathway},~\ref{metastableliquid}, and~\ref{gel}.  (b) Yield at three times.}
\label{yield}
\end{figure*}

\begin{figure*}
\includegraphics[width=\textwidth]{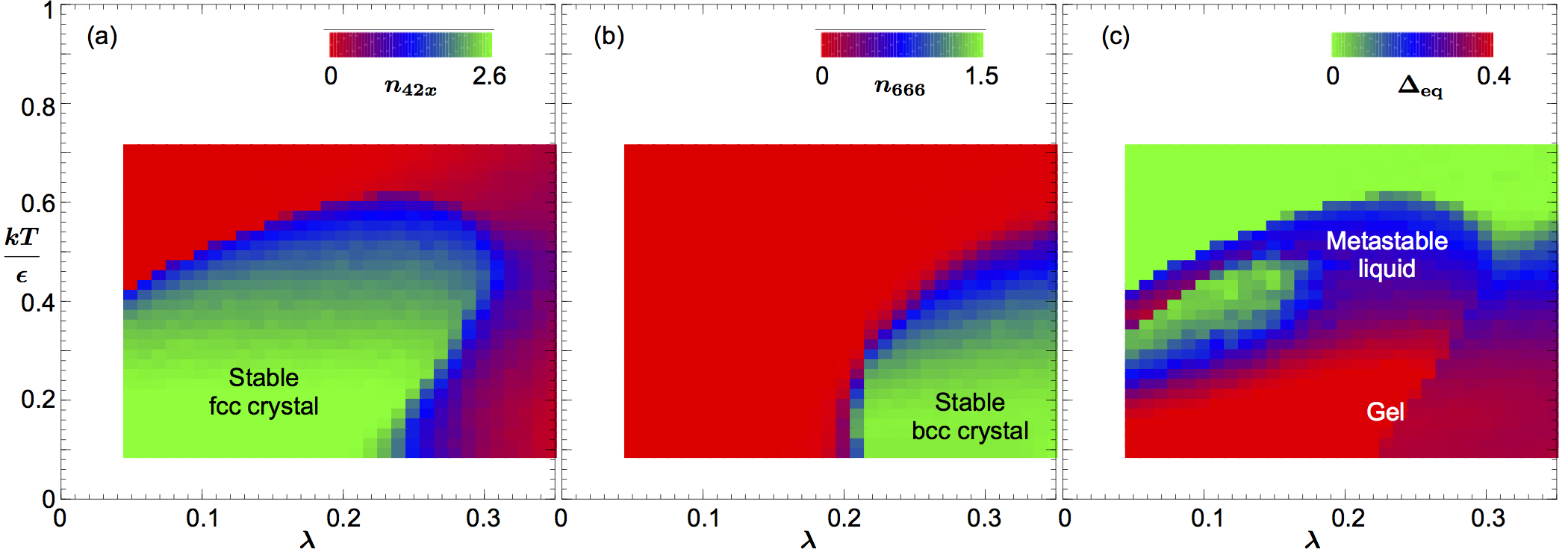}%
\caption{(a) Common neighbor metric $n_{42x}$ indicating fcc crystallinity after simulations of length $t=10^5t_0$ initiated from perfect fcc crystals.  (b) Common neighbor metric $n_{666}$ indicating bcc crystallinity after simulations of length $t=10^5t_0$ initiated from perfect bcc crystals.  (c) Distance-to-equilibrium $\Delta_{\rm eq}$ (defined in Section~\ref{struct}) indicating how close the common neighbor metrics in the dynamic simulations (Fig.~\ref{yield}) get to the closest of the two crystal-initiated simulations.}
\label{thermo}
\end{figure*}

In \f{yield}(a) we show in the temperature-range plane the fcc crystal yield $f_{\rm c}$ seen in dynamic simulations at time $t=10^5 t_0$; panel (b) shows yield also at times $10^3 t_0$ and $10^4 t_0$. High yield (green) is found in a localized region of 
parameter space.  Note that the phase diagrams of  \f{phasediagrams} intersect the diagram of \f{yield} via three vertical lines corresponding to particular values $\lambda=0.18$, $\lambda=0.25$, and $\lambda=0.33$.

A necessary condition for high crystal yield is that the fcc crystal is stable thermodynamically; this 
condition holds
for our 
compact fcc crystals
below the solid green 
curve, marked ``g-c (finite)" in Fig.~\ref{yield}. Note that the dashed green bulk gas-crystal coexistence curve (labeled ``g-c") 
derived from direct coexistence simulations in a slab geometry
lies above the finite-size curve. This difference simply reflects the fact that a finite crystalline cluster with free boundaries can melt within the regime of bulk crystal stability, i.e. can be smaller than the critical cluster size. 
Note also that the solid green curve bends toward small $\lambda$ and small $k_{\rm B}T/\epsilon$ at around $\lambda=0.32,$ $k_{\rm B}T/\epsilon=0.44$.  This bend occurs because the fcc crystal becomes unstable with respect to to a body-centered cubic (bcc) crystal at large $\lambda$ and small $k_{\rm B}T/\epsilon$ (see below).

A second necessary condition for high crystal yield expected from previous work\c{tenWolde1997,lutsko2006ted,Soga1999, Costa2002, Lomakin2003, Fortini2008}
is that a state point must lie within the regime of metastable liquid-gas phase coexistence.
For large interaction ranges $\lambda$ we determined the boundary of liquid stability in bulk from Gibbs ensemble simulations (dashed gray curve in Fig.~\ref{yield}), a technique that eliminates interfacial effects by putting gas and liquid phases in separate boxes that interchange both volume and particles~\cite{Panagiotopoulos1989}.  For interaction ranges smaller than $\lambda=0.15$, we could not accurately determine gas-liquid coexistence because crystallization rapidly occurred within the liquid box.  Instead, we estimated the onset of transient liquid-like structure as the curve below which our dynamic simulations attained a relative number of liquid-like bonds $n_{200}\ge 0.1$ at some point during our dynamic simulations (solid gray curve in Fig.~\ref{yield}; see also Fig.~\ref{diagrams} (g)).  We find that this curve coincides with the bulk liquid curve over the interval of interaction ranges for which both could be calculated, indicating that metastability of the liquid is not strongly influenced by the existence of free boundaries for system sizes on the order of 1000 particles.  
Although it is not clear to what extent the liquids at small interaction range can be considered metastable, comparison of the onset of high crystal yield (green pixels) with the liquid boundary (dashed and solid curves) is consistent with crystallization coinciding with the onset of transient liquid order and/or an extrapolation of the metastable liquid curve.


While our results show that the stability of the crystal and the onset of transient liquid structure are necessary conditions for rapid crystallization, they are clearly not sufficient: large regions of parameter space below the crystal and liquid boundaries appear blue or red in Fig.~\ref{yield}, indicating low crystal yield after $t=10^5t_0$, despite the fact that the systems must eventually assemble into a thermodynamically favored fcc crystal.  The eventual (infinite-time) fate of the system can be inferred from \f{thermo}, in which we plot in panels (a) and (b) the crystal yield that results from simulations initiated from a single fcc or bcc crystal, respectively.  Note that the fcc crystal becomes unstable with respect to the body-centered cubic (bcc) crystal at large $\lambda$ and low $k_{\rm B}T/\epsilon$, because bcc spheres can accommodate 8 nearest neighbors and 8 second-nearest neighbors at these values of $\lambda$, while fcc spheres can only accommodate 12 nearest neighbors. In this region of parameter space the bcc crystal is therefore lower in energy than the fcc crystal. Panel (c) displays a `distance-to-equilibrium' parameter (see Section~\ref{struct}) that summarizes how close dynamic simulations come to equilibrium: anything not shown green corresponds either to a metastable liquid or to an arrested gel.

\begin{figure*}
\includegraphics[width=\textwidth]{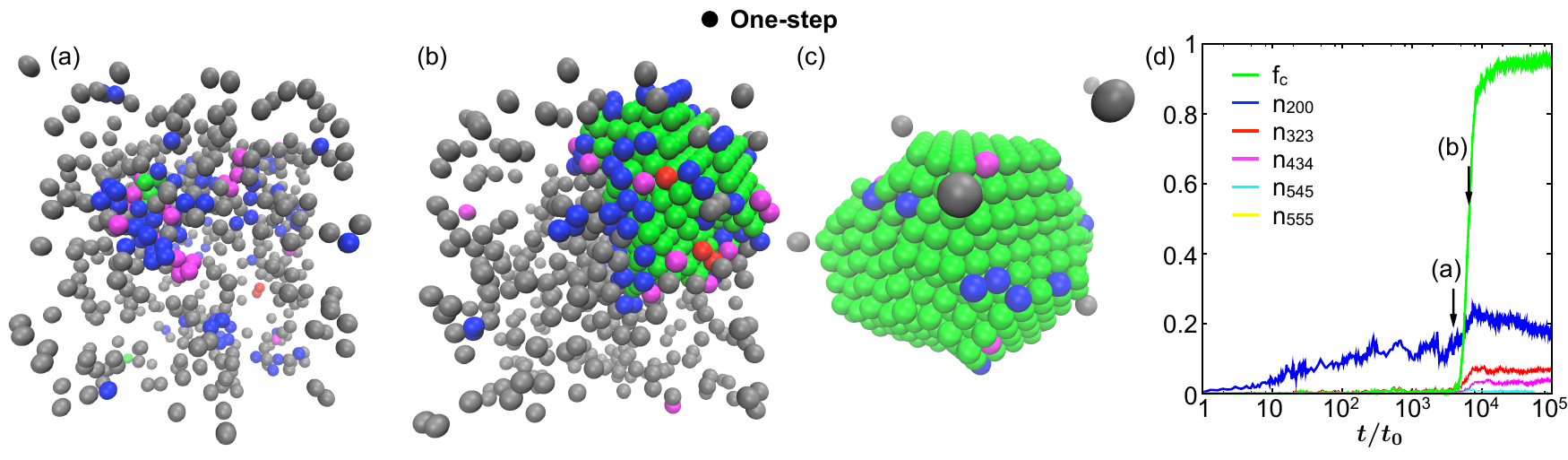}%
\caption{(a-c) Snapshots from a simulation with $\lambda=0.03$ and $k_{\rm B}T/\epsilon=0.3$ that exhibits one-step crystallization (a) before nucleation, (b) after nucleation, and (c) at the end of the simulation ($t=10^5t_0$). (d) Time series of the crystal yield $f_{\rm c}$ and liquid and polytetrahedral common neighbor metrics. Liquid-like environments (blue curve and particles) are seen throughout crystallization, but the crystal that nucleates and grows does not have substantial liquid-like character. Arrows indicate the time points of the snapshots.}
\label{onestep}
\end{figure*}

As we will discuss below, low crystal yield within the thermodynamically stable crystal region is due to one of two kinetic effects, depending on the state point: either nucleation from the liquid is slow, or crystallization is arrested by gelation.  Furthermore, we find that the kinetics of crystallization varies strongly even within the region of high yield: crystallization may proceed in a two-step pathway via a liquid-like intermediate, or it may proceed directly from a relatively homogeneous gas.  The following five subsections discuss the five qualitatively different dynamic regimes encountered when broadly varying the interaction range and strength.



\subsection{Metastable gas}

\begin{figure*}
\includegraphics[width=\textwidth]{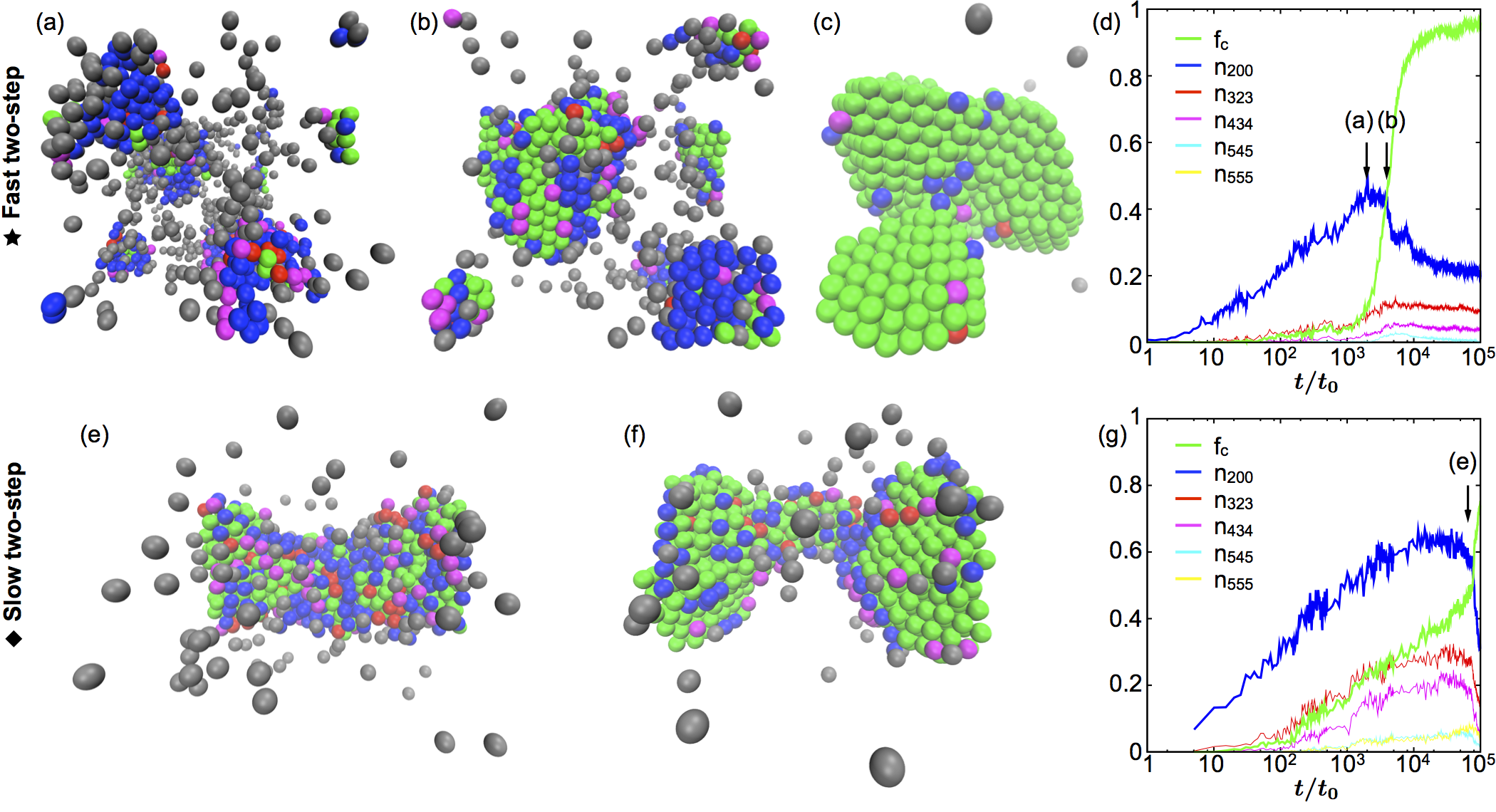}%
\caption{(a-c) Snapshots from a simulation with $\lambda=0.03$ and $k_{\rm B}T/\epsilon=0.28$ exhibiting fast two-step crystallization: (a) before crystal nucleation, (b) after crystal nucleation, and (c) at the end of the simulation ($t=10^5t_0$).  (d) Time series of the crystal yield $f_{\rm c}$ and liquid and polytetrahedral common neighbor metrics.  Arrows indicate the time points of the snapshots.  (e-f) Snapshots from a simulation with $\lambda=0.17$ and $k_{\rm B}T/\epsilon=0.48$ exhibiting slow two-step crystallization: (e) before crystal nucleation and (f) after crystal nucleation, at the end of the simulation ($t=10^5t_0$).  (g) Time series of the crystal yield $f_{\rm c}$ and common neighbor metrics.}
\label{twosteppathway}
\end{figure*}

The black square in \f{yield} lies in the metastable gas regime. Here, as shown in \f{metastable}, the system remains in a gas state, with low values of all common neighbor metrics. In this regime the face-centered cubic (fcc) crystal is the stable state, while the liquid is unstable with respect to the gas.  The metastability of the gas for times up to $t=10^5t_0$ indicates the existence of large free-energy barriers for direct crystal nucleation from the gas.

\subsection{One-step crystallization}

As the temperature decreases below the metastable liquid transition (dashed and solid gray curves in Fig.~\ref{yield}) the crystal yield increases, as indicated by the sharp change from red (low yield) to green (high yield) pixels in Fig.~\ref{yield}.  As shown in Fig.~\ref{onestep} (black circle in Fig.~\ref{yield}), dynamic crystallization pathways near the metastable liquid transition involve crystal nucleation from a fluid with substantial liquid-like fluctuations but no significant gas-liquid phase separation.  Before the sharp increase in crystal fraction $f_c$ at around $6 \times 10^3t_0$ (Fig.~\ref{onestep} (b)), there are substantial fluctuations in the liquid common neighbor metric $n_{200}$, but the average value of $n_{200}$ remains less than $0.2$ (one liquid-like bond per five spheres).  The nucleation event at around $6 \times 10^3t_0$ does not occur at the expense of liquid-like structure, as it would if the crystal nucleated from within a liquid-like droplet.  Instead, the value of $n_{200}$ increases during the nucleation event.  Thus, Fig.~\ref{onestep} illustrates a \textit{one-step} pathway that appears to be facilitated by strong but non-critical density fluctuations.  A similar pathway was identified in Ref.~\cite{klotsa2011predicting}.  As discussed below, the region of parameter space in which we observe a one-step pathway is very narrow, consistent with the fact that it was not found in many other studies.



\subsection{Two-step crystallization}

Beyond this narrow region of one-step nucleation we find a broad range of parameters where crystallization occurs via a two-step pathway, illustrated in Fig.~\ref{twosteppathway} (a-d) and (e-g) (black star and diamond, respectively, in Fig.~\ref{yield}).  First, liquid-like clusters (Fig.~\ref{twosteppathway} (a) and (e)) quickly nucleate, grow, and merge, resulting in an increase in the liquid common neighbor metric $n_{200}$.  Later, crystals (Fig.~\ref{twosteppathway} (b) and (f)) nucleate from within those droplets, resulting in a decrease in $n_{200}$ and an increase in the crystallinity metric $f_{\rm c}$. The nucleation time increases with
increasing range $\lambda$, as can be seen by comparing the common neighbor time series of Fig.~\ref{twosteppathway} (d) and (g). For $\lambda \geq 0.2$ the time for crystal nucleation from the liquid exceeds our simulation time $t=10^5 t_0$.

\begin{figure}
\includegraphics[width=0.5\textwidth]{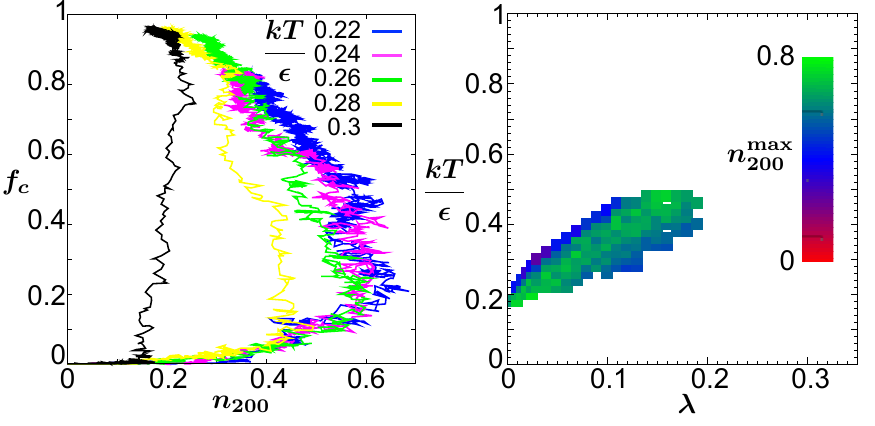}%
\caption{Parametric pathway diagram illustrating the evolution of liquid structure ($n_{200}$) on the horizontal axis and crystalline structure ($f_{\rm c}$) on the vertical axis, for a slice of state points with $\lambda=0.03$ and $0.22\le k_{\rm B}T/\epsilon \le 0.3$. All trajectories show high crystal yield $f_{\rm c}>0.7$ after $t=10^5t_0$.  For most state points ($k_{\rm B}T/\epsilon \le 0.28$) crystallization follows a two-step pathway, where first $n_{200}$ increases and then $n_{200}$ decreases while $f_{\rm c}$ increases.  For $k_{\rm B}T/\epsilon=0.3$ crystallization proceeds via largely a one-step mechanism, with $n_{200}$ remaining low throughout assembly.  (b) Maximum value of $n_{200}$ during assembly as a function of $\lambda$ and $k_{\rm B}T/\epsilon$, restricted to those state points for which $f_{\rm c}>0.7$ after $t=10^5t_0$.  Most points are green, indicating a two-step pathway with large intermediate values of $n_{200}$.  A narrow strip of state points near the extrapolated location of the gas-liquid curve are blue, indicating one-step assembly for which $n_{200}$ is low throughout.}
\label{twostep}
\end{figure}

To illustrate the crossover from one-step to two-step pathway more generally, we show in Fig.~\ref{twostep}(a) a parametric plot of the liquid common neighbor metric $n_{200}$ versus the fractional crystal yield $f_{\rm c}$, for a slice of state points ($\lambda=0.03$ and $0.22\le k_{\rm B}T/\epsilon \le 0.3$) having crystal yields $f_{\rm c}>0.7$ at $t=10^5 t_0$.  Most of these systems ($k_{\rm B}T/\epsilon \le 0.28$, including the example $k_{\rm B}T/\epsilon=0.28$ from Fig.~\ref{twosteppathway}) follow a pronounced two-step pathway. First, $n_{200}$ increases with little increase in $f_{\rm c}$, corresponding to liquid droplet nucleation, growth, and coalescence.  Subsequently, $n_{200}$ decreases and $f_{\rm c}$ increases, corresponding to crystal nucleation (fast or slow) from within the liquid droplet.  In contrast, the example system with $k_{\rm B}T/\epsilon=0.3$ (Fig.~\ref{onestep}) does not exhibit pronounced two-step nucleation; instead, a crystal nucleates directly from the gas. 

In Fig.~\ref{twostep} (b) we show the maximum value of $n_{200}$ along each pathway for which $f_{\rm c}\ge 0.7$ after $t=10^5t_0$.  The narrow strip of values along the top of this region, near the metastable liquid transition, exhibit one-step behavior, with correspondingly low (blue) values of $n_{200}^{\rm max}$.

\subsection{Metastable liquid}

\begin{figure}
\includegraphics[width=0.5\textwidth]{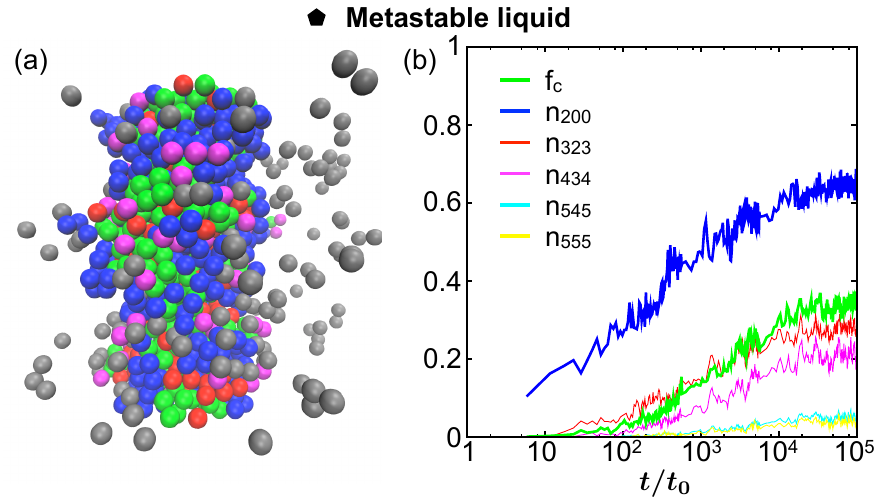}%
\caption{(a) Snapshot from a system with $\lambda=0.2$ and $k_{\rm B}T/\epsilon=0.54$ which remains as a metastable liquid up to time $t=10^5t_0$. (b) Time series of the crystal yield $f_{\rm c}$ and liquid and polytetrahedral common neighbor metrics.}
\label{metastableliquid}
\end{figure}

At larger interaction ranges ($\lambda \geq 0.2$) and for temperatures close to the gas-liquid curve we find that the liquid remains metastable up to times $t=10^5 t_0$. The liquid metric $n_{200}$ is large, and the crystal metric $f_{\rm c}$ increases until it reaches a plateau that persists until the end of the simulation. An example trajectory 
is shown in Fig.~\ref{metastableliquid} (black pentagon in Fig.~\ref{yield}). For larger systems and times that are longer (but still accessible to the corresponding experiments), this region of parameter space may give rise to good crystals.

\subsection{Gelation}

\begin{figure*}
\includegraphics[width=\textwidth]{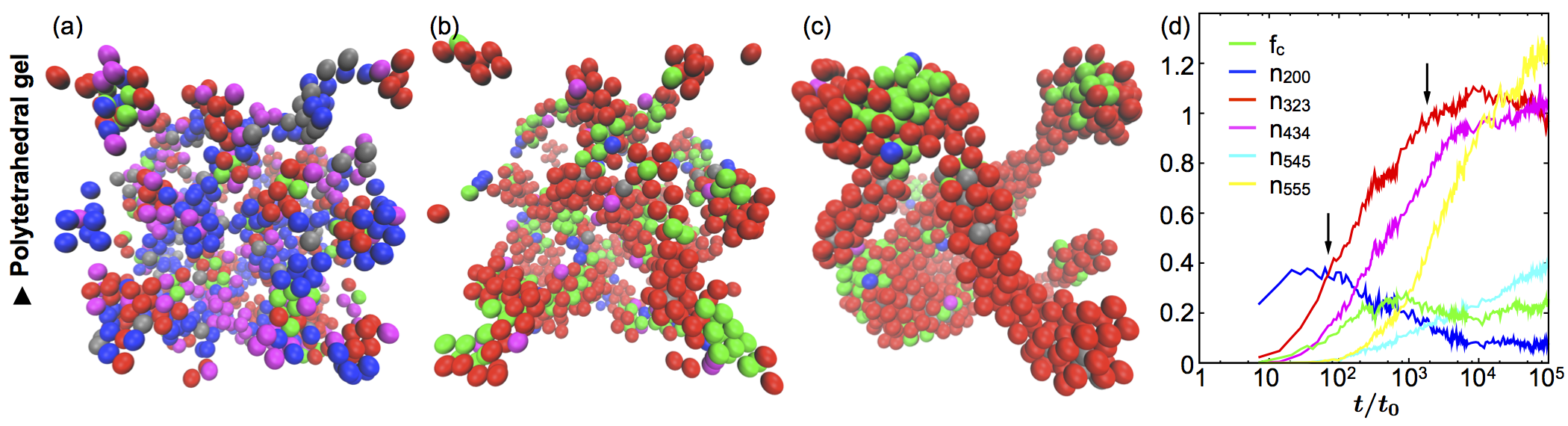}%
\caption{Snapshots from a simulation with $\lambda=0.27$ and $k_{\rm B}T/\epsilon=0.3$ that forms a polytetrahedral gel (a) before gelation, (b) after gelation, and (c) at the end of the simulation ($t=10^5t_0$).  (d) Time series of the crystal yield $f_{\rm c}$ and liquid and polytetrahedral common neighbor metrics.  Arrows indicate the time points of the snapshots.
}
\label{gel}
\end{figure*}

At low temperature, our simulated systems display the fast formation and persistence of polytetrahedral gels~\cite{Campbell2005, Sciortino2005b}, which are bonded networks of face-sharing tetrahedra. As illustrated for small networks in Fig.~\ref{clusters} (b-f), polytetrahedral networks exhibit nonzero values of the common neighbor metrics $n_{212}, n_{323}, n_{434}, n_{545}$, and $n_{555}$.  All but the $n_{212}$ metric do not appear in perfect close-packed crystals; 212 environments appear on the 100 and 110 surfaces
of fcc crystals.  As shown for example in Fig.~\ref{gel} (black triangle in Fig.~\ref{yield}) the metrics $n_{323}, n_{434}, n_{545}$, and/or $n_{555}$ increase quickly 
and remain large up to times $t=10^5t_0$, while metrics characterizing crystals ($f_{\rm c}$) and mobile liquids ($n_{200}$) remain low.

\begin{figure*}
\includegraphics[width=\textwidth]{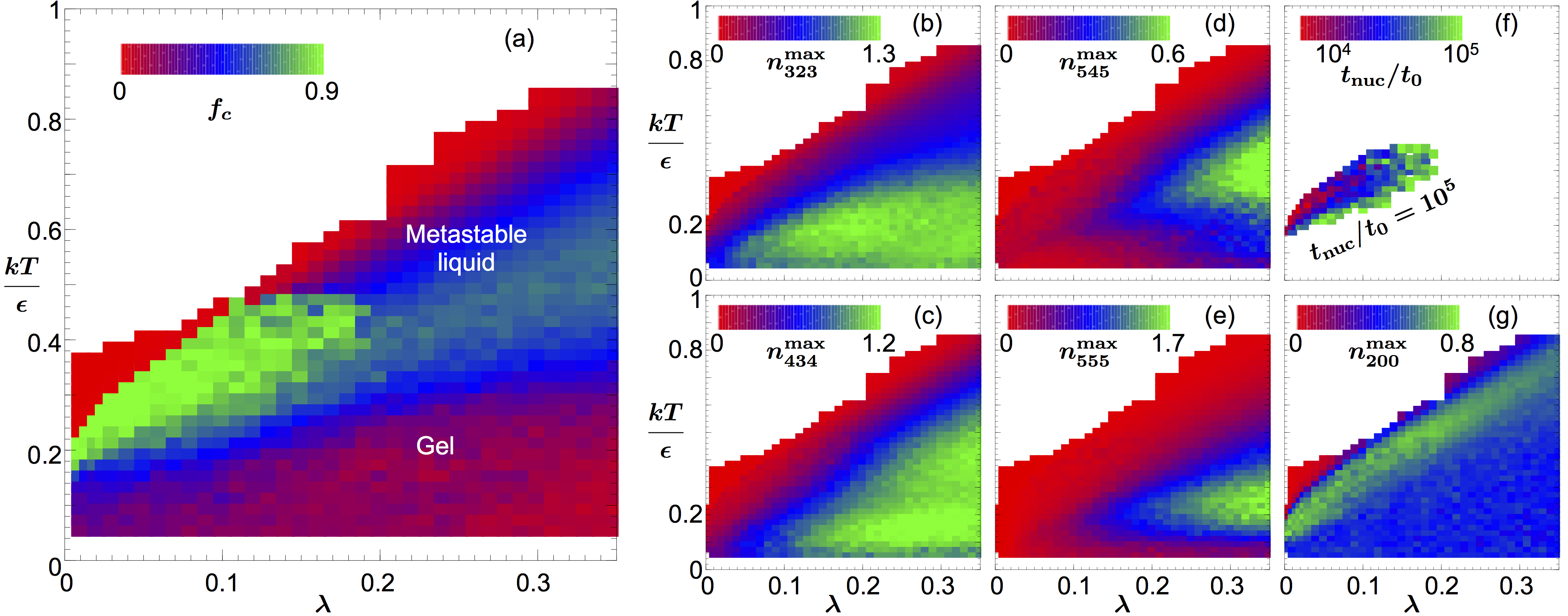}%
\caption{(a) Relative crystal yield $f_{\rm c}$ after $t=10^5t_0$ as a function of interaction range $\lambda$ and temperature $k_{\rm B}T/\epsilon$.  (b-e) Maximum values of the polytetrahedral common neighbor metrics (b) $n_{323},$ (c) $n_{434}$, (d) $n_{545}$, and (e) $n_{555}$ over the course of the simulations, as a function of interaction range $\lambda$ and temperature $k_{\rm B}T/\epsilon$.  The region of parameter space in the lower right of (a) (``gel") where $f_{\rm c}<0.7$ is mostly associated with large values of $n_{323}^{\rm max}, n_{434}^{\rm max}, n_{545}^{\rm max}$, and/or $n_{555}^{\rm max}$ (b-e), except for the region labeled ``metastable liquid".  (f) Nucleation time $t_{\rm nuc}$ (time to first achieve $f_{\rm c}=0.7$) on a logarithmic scale.  (g) Maximum values of the liquid common neighbor metric $n_{200}$.}
\label{diagrams}
\end{figure*}

We find that the decrease in crystal yield at low temperature is accompanied by proliferation of these polytetrahedral networks. This correspondence can be seen across parameter space by referring to Fig.~\ref{diagrams} (a-e), in which we compare (a) the crystal yield $f_{\rm c}$ after $t=10^5t_0$ with (b-e) the maximum value of the polytetrahedral common neighbor metrics $n_{323}, n_{434}, n_{545}$, and $n_{555}$ achieved during the simulations.  Although each common neighbor metric shows somewhat different dependence on $\lambda$ and $k_{\rm B}T/\epsilon$, comparison of Fig.~\ref{diagrams} (a) with Fig.~\ref{diagrams} (b-e) shows that pathways involving large maximum values of the polytetrahedral common neighbor metrics account for most of the region in Fig.~\ref{diagrams} (a) where the close-packed crystals are stable but yield is low.  

The predominance of polytetrahedral common neighbor metrics at low temperature suggests an explanation for the dynamic inaccessibility of the bcc crystal.  Since the region of parameter space where the bcc crystal is stable (Fig.~\ref{thermo} (b)) is contained within the region where the polytetrahedral common neighbor metrics are large (Fig.~\ref{diagrams} (a-e)), there is no region of parameter space where bcc crystallization proceeds efficiently.  Rather, polytetrahedral gelation forestalls bcc crystallization wherever the bcc crystal is thermodynamically stable.

The only large region below the gas-liquid and liquid-crystal curves where neither $f_{\rm c}$ nor the polytetrahedral metrics are large is the region marked ``metastable liquid" in Fig.~\ref{diagrams} (a).  As discussed above, crystal yield is low in this regime because the nucleation time is longer than $10^5 t_0$, not because there is substantial polytetrahedral gelation.  To see that nucleation times extrapolate beyond $10^5 t_0$ in the metastable liquid regime, in Fig.~\ref{diagrams} (f) we plot the time (on a log scale) at which each system first achieves a crystal yield of $f_{\rm c}=0.7$.  Nucleation times increase to the right and top within the high-yield region until they pass beyond the $t=10^5t_0$ window.  Fig.~\ref{diagrams} (g) shows that beyond the high-yield region, the maximum value of $n_{200}$ remains high, indicating that these systems achieve similar levels of liquid structure as in the high-yield region, the main difference being that crystals have not yet nucleated from the liquid after $10^5t_0$.

\section{Conclusions}

Our results confirm that colloidal crystallization displays features common to many examples of self-assembly, in that it happens efficiently in only a small regime or `sweet spot' of parameter space\c{wilber2007reversible,mike,klotsa2011predicting,doi:10.1146/annurev-physchem-040214-121215}. We confirm that efficient crystal nucleation of spherical particles with short-range attractions happens via a two-step pathway throughout most of parameter space.  However, by performing a systematic investigation of colloidal crystallization as a function of interaction strength and range, we have found that largely one-step crystallization from the gas can occur for certain combinations of parameters. Specifically, over a narrow range of interaction strengths and ranges near the extrapolated location of the metastable liquid-gas boundary, we find that crystallization occurs without significant accumulation of liquid-like order within the relatively short time frame of $t=10^5\eta R_0^3/k_{\rm B}T$. 

Our results show that low crystal yield at low temperatures or large interaction strengths is accompanied by particular local bond geometries. Crystal yield is low whenever there are many common neighbor configurations associated with face-sharing tetrahedra. Royall \textit{et al.}~showed that gelation in a colloid-depletant mixture can be explained by formation of overlapping networks of locally-favored states, each defined as a maximally-bonded cluster of size $m \le 13$~\cite{Royall2008, Doye1995}.  For short-range interactions as explored in Ref.~\cite{Royall2008}, most locally-favored states are sections of a 13-particle icosahedron consisting of 20 face-sharing tetrahedra, and are therefore clusters of face-sharing tetrahedra~\cite{Doye1995}.  The only exception is the octahedron (Fig.~\ref{clusters} (g)), which has the same number of spheres (6) and bonds (12) as three face-sharing tetrahedra (Fig.~\ref{clusters} (d)), but is not composed of tetrahedra. The octahedron thus shows a distinct common neighbor signature: instead of exhibiting nonzero values of the polytetrahedral common neighbor metrics, the octahedron exhibits only a nonzero value of the common neighbor metric $n_{200}$ that is prevalent at temperatures above gelation (Fig.~\ref{diagrams} (g)). Our findings are largely consistent with the results of Royall \textit{et al.}, suggesting that key features of nonequilibrium colloidal assemblies can be captured by the square-well model. 

We note also that within the square-well system, gelation can be detected by the study of local bond environments and does require identifying maximally-bonded clusters.  Indeed, our results suggest that there may be some local configurations seen in gels that do not participate in maximally bonded clusters: while 323, 434, and 555 bonds associated with polytetrahedral gelation are found in maximally bonded clusters, 545 bonds are not. Instead, 545 bonds are found in curved, linear polytetrahedral motifs as shown in Fig.~\ref{clusters} (e).  As seen by comparing Fig.~\ref{clusters} (e) to Fig.~\ref{clusters} (f), these motifs do not maximize the number of bonds because they do not close into complete loops of face-sharing tetrahedra.

Finally, our dynamic protocol suggests which combinations of temperature and colloid interaction range will yield best crystallization on the time scale $10^5 t_0$, where $t_0=\eta (2R_0)^3/k_{\rm B}T$. For example, for spherical colloidal particles with radius $R_0=50$ nm, at room temperature ($T=293$ K) in water ($\eta=1.00 \times 10^{-3}$ Pa s), our results predict that crystallization after $t=10^5t_0=25$ s will be best for an attractive interaction of range $0.04R_0=2$ nm and strength $0.32 \, k_{\rm B}T=0.19$ kcal/mol.

\section{Acknowledgements}

This work was performed at the Molecular Foundry, Lawrence Berkeley National Laboratory, supported by the Office of Science, Office of Basic Energy Sciences, of the U.S. Department of Energy under Contract No. DE-AC02--05CH11231. This research used resources of the National Energy Research Scientific Computing Center, a DOE Office of Science User Facility supported by the Office of Science of the U.S. Department of Energy under Contract No. DE-AC02-05CH11231.

\section{Appendix: Virtual-move Monte Carlo algorithm}

To efficiently approximate the overdamped and hydrodynamically coupled dynamics of strongly associating particles in solution, we used the virtual-move Monte Carlo (VMMC) algorithm~\cite{Whitelam2007,Whitelam2011approximating}, specifically the variant described in the appendix of Ref.\cite{Whitelam2009role}.  We parameterized the algorithm to satisfy the Stokes solutions for the translational and rotational diffusion of clusters of hydrodynamic radius $R_{\rm H}$,
\begin{equation}
\begin{array}{l}
D_{\rm t}=\dfrac{k_{\rm B}T}{6\pi\eta R_{\rm H}}, \\
D_{\rm r}=\dfrac{k_{\rm B}T}{8\pi\eta {R_{\rm H}}^3},
\end{array}
\label{Dstokes}
\end{equation}
while allowing as much internal relaxation of each cluster as possible.  Our parameterization allows us to present results across interaction ranges and particles sizes relative to the natural Brownian time unit
\begin{equation}
t_0=\eta (2R_0)^3/k_{\rm B}T,
\label{t0appendix}
\end{equation}
where $\eta$ is the solvent viscosity, $R_0$ is the hard-core radius of the spherical particles, and $k_{\rm B}T$ is the thermal energy.



The VMMC algorithm~\cite{Whitelam2007, Whitelam2009role, Whitelam2011approximating} moves individual particles and groups of particles with attempt and success frequencies designed to (1) preserve the correct equilibrium distribution, (2) ensure that particles move according to gradients in the potential energy, and (3) allow the dependence of diffusion coefficients on cluster size (e.g. Eq. (\ref{Dstokes})) to be controlled.  The algorithm achieves this by proposing individual Monte Carlo moves, self-consistently generating individual or collective moves from the proposed moves, and accepting those moves in a way that satisfies the above three conditions.


In our implementation, trial individual translations are attempted with probability $p_{\rm t}$ by randomly selecting a particle and then attempting to translate it randomly within a ball of radius $\Delta_{\rm t}$.  Because our particles are spherically symmetric, collective rotations cannot be generated from trial rotations about the center of mass of a single particle.
Our trial rotations are attempted with probability $p_{\rm r}=1-p_{\rm t}$ by randomly selecting a particle, randomly selecting a second particle within the interaction range of the first, and then attempting to rotate the second particles by a randomly-chosen angle in the range $(-\Delta_{\rm r}, \Delta_{\rm r})$ around a randomly oriented axis $\hat{n}$ centered at the first particle.

Acceptance rates in the VMMC algorithm consist of three factors: (1) 
a factor built on the Metropolis criterion ensuring that the system relaxes toward equilibrium and particles move according to gradients in the potential energy
(2) a factor ensuring that motions of clusters are not oversampled with respect to the motion of isolated particles, and (3) a factor enforcing a prescribed dependence of diffusion coefficients on cluster size.  
The first factor is generic for any application of the VMMC algorithm.  The second factor is usually used to produce realistic dynamics, but can be omitted when the algorithm is used only to sample an equilibrium distribution.
The third factor has a general form that depends on the scaling of diffusion coefficients $D_{\rm t}$ and $D_{\rm r}$ on hydrodynamic radius $R_{\rm H}$ (e.g.~$D_{\rm t}\propto {R_{\rm H}}^{-1}$ and $D_{\rm r}\propto {R_{\rm H}}^{-3}$ in Eq.~\ref{Dstokes}), but the parameters of the algorithm ($p_{\rm t}$, $\Delta_{\rm t}$, and $\Delta_{\rm r}$) must be tuned to ensure that the prefactor of the diffusion laws match the prescribed values~\cite{Haxton2012slayer}.  Below, we review the VMMC algorithm and discuss our parameter optimization.


\subsection{Review of the VMMC algorithm}

We followed the `symmetrized' version of the VMMC algorithm discussed in Refs.~\cite{Whitelam2009role, Whitelam2011approximating}.  Trial `virtual' single-particle translations and two-particle rotations are generated as discussed above, with probability $p_{\rm t}$ and $p_{\rm r}=p_{\rm t}-1$, respectively.  Trial collective moves are generated by iteratively testing each link between particles inside the moving group and particles outside the moving group, starting with an initial translation or rotation, until no links remain to be tested.  When each link ${ij}$ is tested, the particle outside the moving group ($j$) is \textit{pre-linked} to the moving group with probability
\begin{equation}
p^{\rm link}_{ij}=\mathcal{I}_{ij}\textrm{max}\left(0, 1-\exp\left(\beta(U_{ij}-U_{i^{\prime}j})\right)\right),
\label{plink}
\end{equation}
where $\mathcal{I}_{ij}=1$ if particles $i$ and $j$ are within their mutual interaction range and $\mathcal{I}_{ij}=0$ otherwise, $U_{ij}$ is the initial interaction energy, and $U_{i^{\prime}j}$ is the interaction energy when particle $i$ executes a virtual move but particle $j$ does not. After making a virtual move particle $i$ is returned to its initial coordinates. If particle $j$ is pre-linked to the moving group, then the probability of the reverse move (translation or rotation in the opposite direction) is calculated,
\begin{equation}
p^{\rm reverse}_{ij}=\textrm{max}\left(0, 1-\exp\left(\beta(U_{ij}-U_{i^{\prime\prime}j})\right)\right),
\label{preverse}
\end{equation}
where $U_{i^{\prime\prime}j}$ is the interaction energy when particle $i$ does the reverse virtual move and particle $j$ stays still. After the reverse virtual move $i$ is restored to its initial position. Then, the pre-linked particle $j$ is {\em linked} (admitted) to the moving group with probability
\begin{equation}
p^{\rm link}_{ij}=\textrm{min}\left(1, \dfrac{p^{\rm reverse}_{ij}}{p^{\rm link}_{ij}}\right);
\label{plink}
\end{equation}
otherwise, the link is marked as \textit{frustrated}.  This procedure is continued until it converges on a final moving group, with all links having been tested between particles inside and outside the group.

Since the algorithm will move the entire group by the initial translation or rotation, the acceptance of collective moves must be scaled by a factor of $1/N$ to prevent over-sampling the motion of particles that are in large clusters.  This is achieved by rejecting moves \textit{in situ} when the number of particles in a moving group grows larger than $1/x$ (translations) or $2/x$ (rotations), where $x$ is a random number on the interval $(0,1]$ chosen at the beginning of each Monte Carlo step.  The factor 2 appears because only rotations of groups of at least two particles are explicitly simulated for spherical particles; rotations of individual particles can be assumed to occur at any arbitrary rate while having no effect on the center of mass motion of the particles.

An additional scaling of acceptance probabilities can be applied to control the dependence of diffusion on cluster size.  
The Stokes scalings (Eq.~\ref{Dstokes}) can be perfectly enforced in the limit where particles are tightly bound to each other within well-defined clusters and do not interact outside of these clusters.  In this limit, any trial Monte Carlo move results in the recruitment of the entire cluster into a moving group, followed by the translation or rotation of the entire cluster.  The VMMC algorithm can enforce Eq.~\ref{Dstokes} in this limit by rejecting moves \textit{in situ} when the hydrodynamic radius $R_{\rm H}$ of the moving group exceeds $R_{\rm min}/y$ (translations) or $R_{\rm min}/y^{3}$ (rotations), where $R_{\rm min}$ is the minimum possible hydrodynamic radius (see below) and $y$ is another random number on the interval $(0,1]$.  Following previous implementations of the VMMC algorithm~\cite{Whitelam2007} we estimate the hydrodynamic radius $R_{\rm H}$ of a group $\mathcal{G}$ as a generalization of the radius of gyration,
\begin{equation}
{R_{\rm H}}^2\equiv 10\langle |({\vec{r}}-{\vec{r}}_{\rm center})\times \hat{n}|^2\rangle_{{\vec{r}}\in\mathcal{G}},
\label{hyd}
\end{equation}
where ${\vec{r}}_{\rm center}$ is the group's center of mass (center of rotation) and $\hat{n}$ is the direction of the translation (axis of rotation) for translations (rotations). This factor is the same for for translations that occur in opposite directions, and for rotations that occur with opposite sense, as is required for detailed balance. We take $\vec{r}\in\mathcal G$ to include all points within the hard cores of the particles.  The minimum hydrodynamic radius for both translations and rotations (single-sphere translations or effective two-sphere rotations about an axis $\hat{n}$ parallel to the separation vector between the particles) is 
the physical sphere radius $R_0$ (it would be $R_0/\sqrt{10}$ without the factor of 10 that appears in \eqq{hyd}).

Once a moving group has been generated that is not rejected \textit{in situ} due to its number of particles or hydrodynamic radius, two additional factors contribute to its acceptance probability.  First, the move is rejected if there are any frustrated links between particles inside and outside the moving group.  Second, moves that remain valid are accepted with probability
\begin{equation}
W_{\rm acc}=\textrm{min}\left(1, \prod_{\langle ij \rangle_{0\leftrightarrow \textrm{p}}}\exp\left(-\beta\left(U_{i^{\prime}j}-U_{ij}\right)\right)\right),
\label{accept}
\end{equation}
where the product runs over all pairs of particles ($i$ in the moving group and $j$ outside it) that are non-interacting before the move and have positive pair energy after the move, or vice versa.  Together, these factors ensure that the system satisfies 
superdetailed balance, a condition that implies detailed balance~\cite{Frenkel2004}.  
For square well spheres that have only zero, negative, or infinite positive interaction, Eq.~\ref{accept} reduces to a rejection if the move results in any hard-core overlap,
\begin{equation}
W_{\rm acc}=\prod_{i\in\mathcal{G}, j\notin\mathcal{G}}\theta(r_{i^{\prime}j}-2R_0),
\end{equation}
where $\theta$ is the heaviside step function.

For rotations of moving groups large enough to interact with their periodic images we imposed an additional rejection if a move resulted in a hard core overlap with a periodic image.  Such overlaps occurred only for gels.

\subsection{Parameter optimization}

The prescribed dependence of diffusion coefficients with cluster size (Eq.~\ref{Dstokes}) is derived for the VMMC algorithm in the limit of vanishingly narrow potential energy wells.  In this limit, trial moves always take particles out these wells.  If the wells are deep relative to $k_{\rm B}T$ and the clusters are isolated, this causes the algorithm to recruit the entire cluster into the moving group, always accept the move, and thus generate diffusion coefficients dictated by the \textit{in situ} rejection of collective moves as a function of hydrodynamic radius.

In practice, potential energy wells are not vanishingly narrow.  Properly modeling the motion of such systems requires (1) allowing degrees of freedom internal to the clusters to relax and (2) ensuring that the prescribed diffusion laws are obeyed even when whole-cluster moves are not always generated.  We sought to satisfy these conditions by choosing algorithm parameters that allow clusters to internally relax as much as possible without violating the Stokes scaling.
We achieved this by selecting the smallest values of $\Delta_{\rm t}$ and $\Delta_{\rm r}$ that resulted in the Stokes solutions for a test set of tetrahedral clusters of size 1 to 120 (8 spheres along each edge) 
for a range of interaction ranges $\lambda$
and $T\rightarrow 0$ (infinite square well attractive interaction).
We determined the optimal algorithm parameters in four steps.

\begin{figure}
\includegraphics[width=0.5\textwidth]{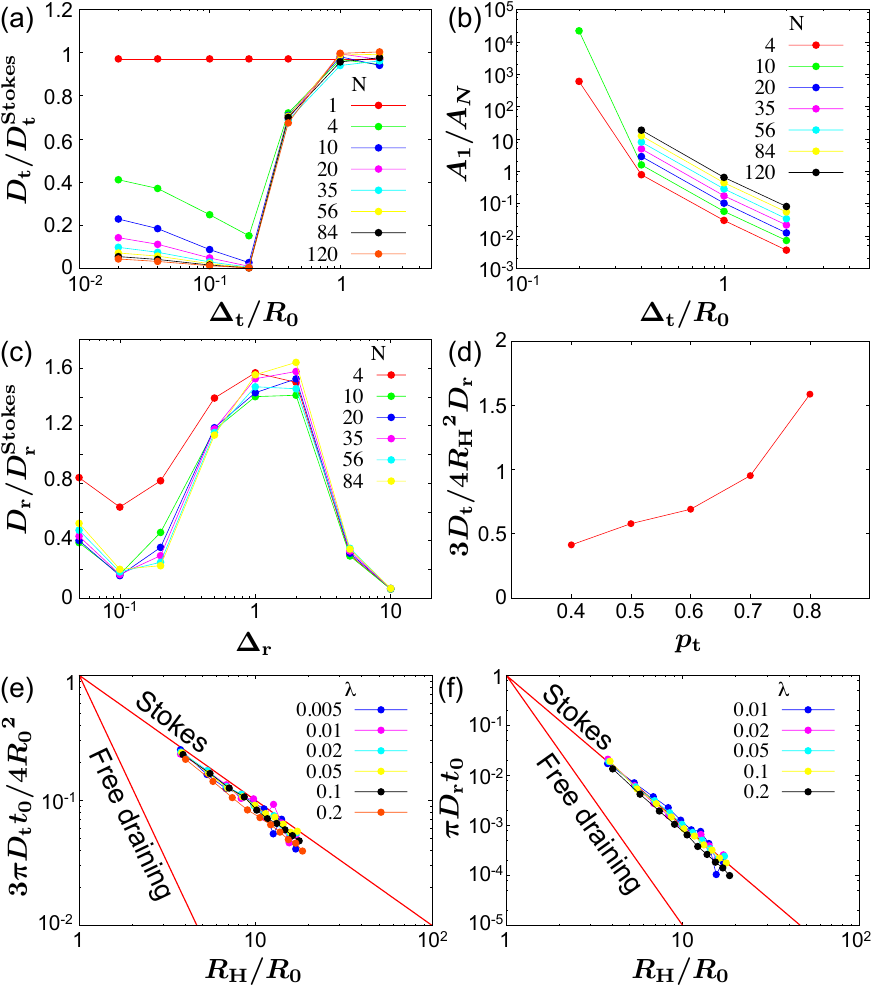}%
\caption{(a) Ratio of translational diffusion coefficient to the value predicted by the Stokes solution (Eq.~\ref{DStokest}) for translation-only VMMC simulations of tetrahedral clusters composed of spheres with infinite square well attractive interactions of range $\lambda=0.11$.  (b) Ratio of accepted single-sphere moves to whole-cluster moves for the same set of simulations.  (c) Ratio of rotational diffusion coefficient to the value predicted by the Stokes solution (Eq.~\ref{DStokesr}) for rotation-only VMMC simulations of the same tetrahedral clusters.  (d) Ratio of translational to rotational diffusion coefficients for a four-sphere tetrahedron with infinite square well attractive interactions as a function of the attempt probability for translations.  The ratio is normalized such that the Stokes solution corresponds to a value of 1 (see Eq.~\ref{Dratio}).  (e) Translational and (f) rotational diffusion coefficients vs hydrodynamic radius for simulations of tetrahedral clusters of various sizes composed of infinite square well attractive spheres with various interaction ranges $\lambda$ (legend), using the full translating and rotating VMMC algorithm with parameters $\Delta_{\rm t}=4\lambda R_0$, $\Delta_{\rm r}=1$, and $p_{\rm t}=0.4(R_0\Delta_r/\Delta_t)^2$.  For comparison, the Stokes solutions ($D_{\rm t}\propto {R_{\rm H}}^{-1}$ and $D_{\rm r}\propto {R_{\rm H}}^{-3}$) and free draining solutions ($D_{\rm t}\propto {R_{\rm H}}^{-3}$ and $D_{\rm r}\propto {R_{\rm H}}^{-5}$) are shown as lines.}
\label{diffusionappendix}
\end{figure}

First, we recorded the translational diffusion coefficient $D_{\rm t}$ from translation-only VMMC simulations of isolated tetrahedral clusters with range $\lambda=0.11$ and compared it to the predicted value from Eq.~\ref{Dstokes},
\begin{equation}
D_{\rm t}^{\rm Stokes}=D_{\rm t}^0\dfrac{R_0}{R_{\rm H}},
\label{DStokest}
\end{equation}
where
\begin{equation}
D_{\rm t}^0=\dfrac{{\Delta_{\rm t}}^2}{10t_{\rm cycle}}
\label{DStokest0}
\end{equation}
is the translational diffusion coefficient for a single sphere translating with Monte Carlo dynamics.  As shown in Fig.~\ref{diffusionappendix} (a), we found that for a range of tetrahedral cluster sizes $D_{\rm t} \ll D_{\rm t}^{\rm Stokes}$ for $\Delta_{\rm t} \le 0.2 R_0$, $D_{\rm t}\simeq 0.7 D_{\rm t}^{\rm Stokes}$ for $\Delta_{\rm t}=0.4 R_0$, and $D_{\rm t} \simeq D_{\rm t}^{\rm Stokes}$ for $\Delta_{\rm t} \ge R_0$.
Although the translational diffusion coefficient is not quite saturated at the Stokes limit at $\Delta_{\rm t}=0.4 R_0$, we chose to parameterize $\Delta_{\rm t}$ near this value because this balances a diffusion coefficient near the Stokes limit with ample internal relaxation: as shown in Fig.~\ref{diffusionappendix} (b), more single-particle than whole-tetrahedron moves are accepted for $\Delta_{\rm t}=0.4 R_0$, while the reverse is true for $\Delta_{\rm t}=R_0$.

Second, we recorded the rotational diffusion coefficient $D_{\rm r}$ from rotation-only VMMC simulations of the same tetrahedral clusters and compared it to the predicted value from Eq.~\ref{Dstokes},
\begin{equation}
D_{\rm r}^{\rm Stokes}=D_{\rm r}^0\left(\dfrac{R_0}{R_{\rm H}}\right)^3,
\label{DStokesr}
\end{equation}
where
\begin{equation}
D_{\rm r}^0=\dfrac{\theta_{\rm ms}(\Delta_{\rm r}){\Delta_{\rm r}}^2}{6t_{\rm cycle}}
\label{DStokesr0}
\end{equation}
is the rotational diffusion coefficient for a single sphere rotating with Monte Carlo dynamics.  In Eq.~\ref{DStokesr0} $\theta_{\rm ms}(\Delta_{\rm r})$ is the mean-squared rotation angle (about a fixed arbitrary axis centered at the center of mass) that a sphere would experience if we applied our Stokes-scaled VMMC algorithm to volume elements within the sphere.  This factor is necessary for rotations because, unlike for translations, the hydrodynamic radius (Eq.~\ref{hyd}) depends on the center and axis of rotation.  Defining $\theta(\Delta_{\rm r}, \hat{n}, \vec{r})$ as the center-of-mass rotation angle for a rotation by $\Delta_{\rm r}$ around $\hat{n}$ centered at $\vec{r}$, we find
\begin{equation}
\theta_{\rm ms}(\Delta_{\rm r})=\dfrac{1}{{\Delta_{\rm r}}^2}\left\langle\left(\theta(\Delta_{\rm r}, \hat{n}, \vec{r}_{\rm center})\right)^2\left(\dfrac{R_0}{R_{\rm H}(\vec{r}_{\rm c}, \hat{n})}\right)^3\right\rangle_{\hat{n}, \left|\vec{r}_{\rm c}\right|<R_0},
\label{c}
\end{equation}
where
\begin{equation}
R_{\rm H}(\vec{r}_{\rm c}, \hat{n})= \left(\dfrac{15}{2\pi R_0^3}\int_{\vec{r}<R_0} |(\vec{r}-\vec{r}_{\rm c})\times \hat{n}|^2\right)^{1/2}
\end{equation}
is the center- and axis-dependent hydrodynamic radius.
In the limit $\Delta_{\rm r}\rightarrow 0$, we numerically calculated $\theta_{\rm ms}(\Delta_{\rm r})\rightarrow \theta_{\rm ms}^0\simeq 0.14$.
As shown in Fig.~\ref{diffusionappendix} (c), we find that for a range of tetrahedral cluster sizes $D_{\rm r}$ saturates from below near a value $1.5D_{\rm r}^{\rm Stokes}$ for $\Delta_{\rm r}\ge 1$.  ($D_{\rm r}$ drops again for $\Delta_{r}\ge 5$ due to the inability to resolve rotations when individual rotations exceed $\pi$ radians.)  We chose to parameterize $\Delta_{\rm r}=1$ to ensure that rotations are saturated at the large-$\Delta_{\rm r}$ limit, allowing translations to accommodate internal relaxation.

Third, we adjusted the attempt probability for translation and rotation, $p_{\rm t}$ and $p_{\rm r}=1-p_{\rm t}$, to correct for the numerical discrepancies between the predicted and measured diffusion coefficients for tetrahedral clusters (the factors 0.6 for translations and 1.5 for rotations discussed above).  We achieved this by performing VMMC simulations of a tetrahedra of four spheres with various translation attempt probabilities $p_{\rm t}$ and comparing the relative diffusion coefficients to the expected relationship (see Eq.~\ref{Dstokes})
\begin{equation}
\dfrac{D_{\rm t}}{D_{\rm r}}=\dfrac{4}{3}{R_{\rm H}}^2.
\label{Dratio}
\end{equation}
If the prefactors for the diffusion coefficients followed Eq.~\ref{DStokest0} and Eq.~\ref{DStokesr0}, Eq.~\ref{Dratio} would be satisfied for
\begin{equation}
\dfrac{p_{\rm t}}{p_{\rm r}}=\dfrac{20}{9}\theta_{\rm ms}(\Delta_{\rm r})\left(\dfrac{R_0\Delta_{\rm r}}{\Delta_{\rm t}}\right)^2=c\left(\dfrac{R_0\Delta_{\rm r}}{\Delta_{\rm t}}\right)^2.
\label{pratio}
\end{equation}
Inserting the $\Delta_{\rm r}\rightarrow 0$ limit, $\theta_{\rm ms}(\Delta_{\rm r})\simeq 0.14$, into Eq.~\ref{pratio}, we find $c\simeq 0.31$ and $p_{\rm t}\simeq 0.6$.  Instead, Fig.~\ref{diffusionappendix} (d) shows that $p_{\rm t}=0.7$ (corresponding to $c=0.4$) results in better agreement with Eq.~\ref{Dratio}.  We therefore fix $p_{\rm t}$ via Eq.~\ref{pratio} with $c=0.4$.

Finally, we showed that fixing the parameters as above results in diffusion coefficients agreeing with the Stokes solutions (Eq.~\ref{Dstokes}) for tetrahedra of various sizes and with various interaction ranges.  We fixed the translational step size $\Delta_{\rm t}=4\lambda R_0$ to be proportional to the interaction range to ensure that the ratio of single-particle to whole-cluster moves be consistent across interaction ranges.  We fixed $\Delta_{\rm r}=1$ and $p_{\rm t}=0.4(R_0\Delta_r/\Delta_t)^2$ as described above.  Together, this parameterization fixes the time per Monte Carlo cycle,
\begin{equation}
t_{\rm cycle}=\dfrac{6}{5}\pi p_{\rm t}\lambda^2 t_0,
\end{equation}
where $t_0$ is the natural Brownian time scale (Eq.~\ref{t0appendix}).
Fig.~\ref{diffusionappendix} (e) and (f) show that $D_{\rm t}$ and $D_{\rm r}$ follow the Stokes solutions (Eq.~\ref{Dstokes}) over a broad range of cluster sizes and interaction ranges, in stark contrast to single-particle Monte Carlo simulations, which follow the much more strongly size-dependent free draining solutions $D_{\rm t}\propto {R_{\rm H}}^{-3}$ and $D_{\rm r}\propto {R_{\rm H}}^{-5}$ for small step sizes.

\footnotesize{
\bibliography{squarewellium.bbl} 
\bibliographystyle{rsc} 
}
\normalsize

\end{document}